\documentclass[reprint, amsmath,amssymb, aps, prb, hyperref]{revtex4-1}
\usepackage{graphicx,xcolor}
\usepackage{dcolumn}
\usepackage{bm}
\newcommand{\refa}[1]{\textcolor{black}{#1}}
\newcommand{\refb}[1]{\textcolor{black}{#1}}

\begin{document}

\title{Selective damping of plasmons in coupled two-dimensional systems by Coulomb drag}

\author{Ilya Safonov}
 \email{safonov.iv@phystech.edu}
  \affiliation{
  Moscow Institute of Physics and Technology, Dolgoprudny 141700, Russia}
 \affiliation{Programmable Functional Materials Lab, Center for Neurophysics and Neuromorphic Technologies, Moscow 127495}
 \author{Aleksandr S. Petrov}
\author{Dmitry Svintsov}
\affiliation{
Moscow Institute of Physics and Technology, Dolgoprudny 141700, Russia}

\date{\today}

\begin{abstract}
		
 The Coulomb drag is a many-body effect observed in proximized low-dimensional systems. It appears as emergence of voltage in one of them upon passage of bias current in another. The magnitude of drag voltage can be strongly affected by exchange of plasmonic excitations between the layers; however, the reverse effect of Coulomb drag on properties of plasmons has not  been studied. Here, we study the plasmon spectra and damping in parallel two-dimensional systems in the presence of Coulomb drag. We find that Coulomb drag leads to selective damping of one of the two fundamental plasma modes of a coupled bilayer. For identical electron doping of both layers, the drag suppresses the acoustic plasma mode; while for symmetric electron-hole doping of the coupled pair, the drag suppresses the optical plasma mode. The selective damping can be observed both for propagating modes in extended bilayers and for localized plasmons in bilayers confined by source and drain contacts. The discussed effect may provide access to the strength of Coulomb interaction in 2d electron systems from various optical and microwave scattering experiments.
 
		
	\end{abstract}
	
	\maketitle
	
	\section{\label{sec:level1} Introduction }
	
	The Coulomb drag effect is a prominent manifestation of the long-range Coulomb interactions~\cite{narozhny2016coulomb}. This effect is usually observed in two proximized conducting layers electrically isolated from each other. When a direct current $I_{drive}$ is applied to one of the systems ("active"), charge carriers in that layer induce Coulomb-mediated friction and drag the carriers in the "passive" layer. The latter eventually hosts a non-zero voltage drop $V_{drag}$. The natural combination of these quantities $V_{drag}/I_{drive}$ is called drag resistance.
	
	The drag resistance depends on various system parameters (e.g., temperature, distance between the layers, magnetic field, etc.) and these dependences incorporate valuable information about the many-body charge carrier interactions. Thus, if both layers are in a Fermi-liquid state, the drag resistance rises quadratically with the temperature $T$, repeating the $T$-dependence of quasiparticle scattering rate. Any deviations from this rule would mean either a non-Fermi-liquid behavior~\cite{debray2002coulomb,laroche20141d,pillarisetty2005coulomb,eisenstein2014exciton} or an additional interaction mechanism~\cite{gramila1993evidence,hill1997correlation,solomon1991mutual,boev2019coulomb}. One of these mechanisms represents the interlayer exchange with collective charge density oscillations, or plasmons~\cite{flensberg1994coulomb,flensberg1995plasmon}. It was argued that although plasmonic dispersions lie outside of the particle-hole continuum at $T=0$, at finite temperatures $\sim0.5$ of Fermi energy thermally excited quasiparticles and plasmons may interact. This should lead to the dramatic increase in the drag resistivity. This prediction was justified in later experiments~\cite{hill1997correlation,noh1998many} for weakly-interacting 2DEGs. Further theoretical research~
\cite{badalyan2007exchange,principi2012plasmons,chen2015boltzmann,kainaris2017coulomb,fandan2019effect,zverevich2023transport} proposed other regimes when plasmon contribution to Coulomb drag might be crucial, and these findings have yet to be experimentally confirmed.
	
The reverse situation, where the many-body quasiparticle interactions affect the properties of plasmon modes, is under active study currently. In particular, strong electron-electron scattering may lead to the hydrodynamic regime of electron flow~\cite{bandurin2016negative,krishna2017superballistic,aharon2022direct}. Already at the level of 'ideal' electron fluid, the plasmon velocity in the hydrodynamic regime should differ from that in the collisionless regime~\cite{Svintsov_crossover,Lucas_plasmons_hydro,hofmann2022collective}. Experimental evidence for this difference were obtained in recent scanning near-field experiments on graphene~\cite{Ruiz2023a} at relatively long plasmon wave lengths (0.5$\mu$m and longer). At shorter wave lengths, the ideal hydrodynamics is no more applicable, and extra 'viscous' corrections to the plasmon damping~\cite{semenyakin2018alternating,petrov2022viscosity,hasdeo2023coulomb} should be evaluated and added to the ordinary damping due to impurities and phonons. 

The enlisted electron interaction effects on plasmons become even richer in the magnetic fields. Despite the absence of fundamental cyclotron resonance renormalization (Kohn theorem~\cite{Kohn}) in non-relativistic systems, the interaction effects on magnetoplasmons may be present in graphene~\cite{Levitov_interactions_and_plasmons} or in the presence of electromagnetic retardation~\cite{Zabolotnykh_Electrical_CR,Andreev_CR_renormalization}. They also may be profound for high-order cyclotron resonances and emanating Bernstein modes~\cite{bandurin2022cyclotron,Alekseev_viscous_resonance,kapralov2022ballistic,afanasiev2023new}. \refa{Importantly, the mentioned effects of electron viscosity on plasmon frequency and damping are proportional to the wave vector squared, i.e. they become pronounced only at relatively short wavelengths. This contrasts to the damping due to impurities and phonons, which is wave-vector independent and thus finite at $q\rightarrow 0$.}

\begin{figure*} \label{fig:Loss_function}
	\centering
	\includegraphics[width=\textwidth]{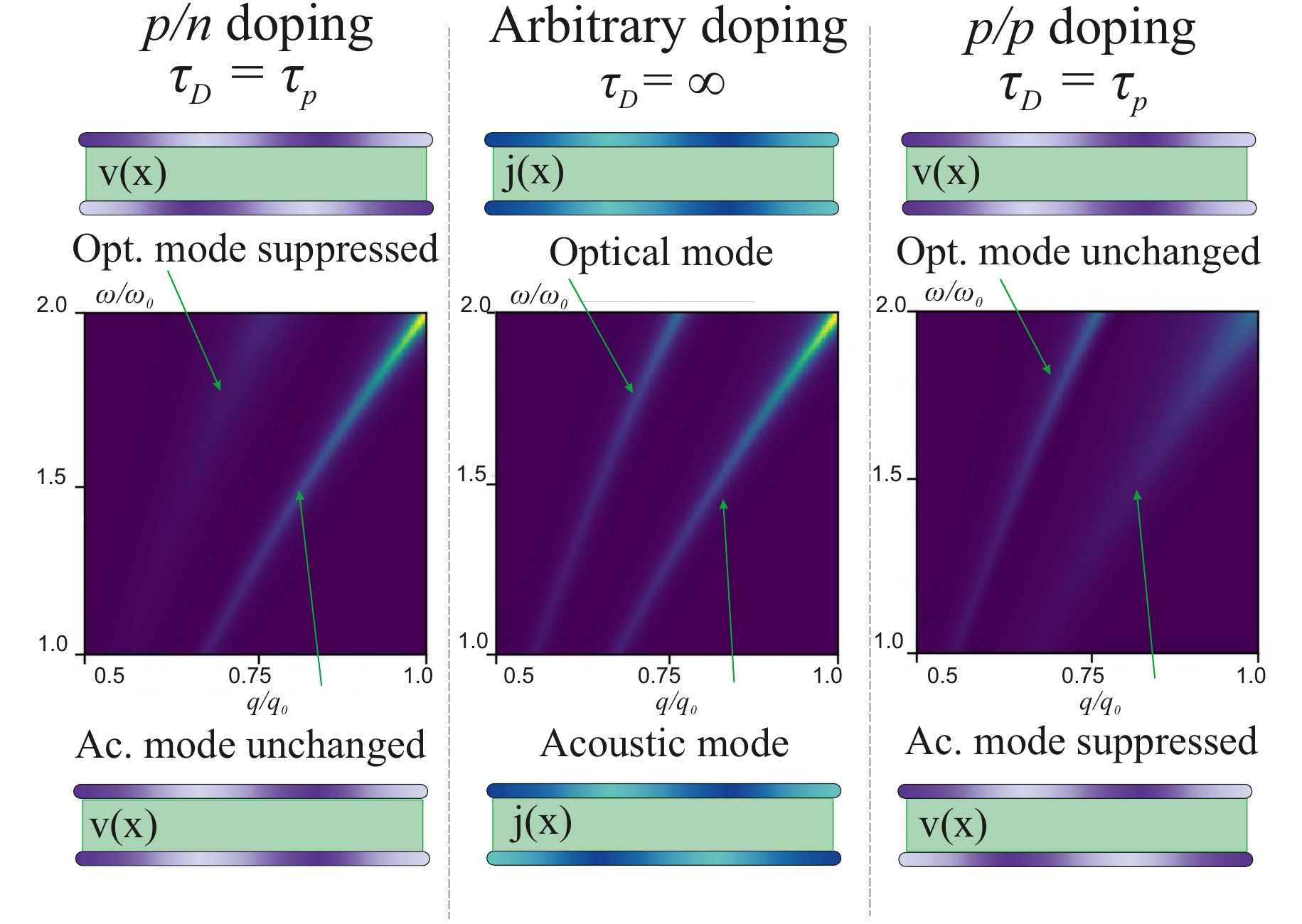}
	\caption{Selective influence of Coulomb drag on plasmon resonances in coupled 2DESs with equal conductivities. \textit{Central column:} loss function and spatial distributions of \textit{current density} for acoustic and optical modes in the absence of Coulomb drag. For the definition of the loss function, see Section~\ref{secII}. Crucially, the current density distribution does not depend on the carrier type in neither of the 2DESs. \textit{Left column:} Schematic spatial distributions of carrier \textit{velocity} for acoustic and optical modes for $p$/$n$ layer doping in the presence of Coulomb drag. The width of the acoustic mode is unaffected by the drag as its velocity distribution is symmetric, leading to no drag friction. The optical mode is suppressed due to asymmetric velocity distribution. \textit{Right column:} the same as in the left column, but for $p$/$p$ doping of the layers. Switching the carrier type in one of the layers switches the velocity distribution and allows to selectively attenuate either of the modes. \textit{Parameters of calculation:} $q_0=2/d$, $\omega_0 = 7\cdot 10^{13}\,$rad/s, $\tau_p^{-1} = 3.5\cdot10^{12}\,$s, $d=2\,$nm.}
\end{figure*}
	
\refa{In this paper, we show that Coulomb drag in double-layer resonators leads to the selective damping of plasmon modes in such a way that the damping rate stays finite at long wavelengths $q \rightarrow 0$. This contrasts to the viscous damping rate being roughly equal to $\nu q^2$, where $\nu$ is the kinematic viscosity. In other words, the effect of Coulomb drag damping is ''more local'' than the effect of viscous damping, though both effects appear already at the level of hydrodynamic equations applicable at relatively long wavelengths.}

The drag resistance, as any other friction mechanism, shortens plasmon lifetimes in DLRs. Remarkably, this shortening {\it selectively} acts on one of the two fundamental plasmon modes (acoustic or optical), depending on carrier polarity. The sensitivity of the drag-induced damping to mutual carrier polarity stems from the fact that Coulomb friction relaxes the \textit{carrier velocity} difference between the layers, not the current contrast~\footnote{Indeed, if the carriers move with equal velocities, we can switch to the inertial reference frame where the particles are at rest, with obviously no drag force acting on them.}. A related effect of selective plasmon damping was predicted for electron-hole plasma in graphene~\cite{svintsov2012hydrodynamic}, yet these two charged components were inseparable in space.
 
 The key result of our study is anticipated in Fig.~\ref{fig:Loss_function}, where the schematic plasmon dispersions in the absence (middle column) and in the presence (side columns) of Coulomb drag are presented. This system supports two types of modes~\cite{sarma1981collective,kainth1999angle,morozov2018giant}: lower in-phase acoustic mode with antisymmetric current distribution and upper out-of-phase optical mode with symmetric current distribution (see Fig. \ref{fig:Loss_function}, central column). Importantly, these distributions are insensitive to carrier types in the layers and depend only on the conductivities of the 2DESs. If such a system hosts Coulomb drag friction, one of the modes acquires additional damping. For example, if both layers are doped either with holes or with electrons (right part of Fig.~\ref{fig:Loss_function}), then the optical mode is unaffected by the drag friction as carrier \textit{velocity} distributions in the layers are identical, while the acoustic mode with antisymmetrical velocity distribution acquires additional linewidth and is greatly suppressed. The reverse situation takes place if the layers are doped with carriers of opposite sign, see left column of Fig.~\ref{fig:Loss_function}.

 \section{Coulomb drag and plasmons in infinite Double-Layer Resonators}\label{secII}
	
	In this Section we construct a theoretical model for the description of plasmon waves in an infinitely long DLR in the presence of Coulomb drag.
	
\refb{	We describe the electron kinetics with account for the Coulomb drag effect using the usual Drude-like theory with the presence of mutual friction between charge carriers in different layers of a two-layer structure \cite{pogrebinskii1977mutual,svintsov2012hydrodynamic,lifshitz1981physical}
	\begin{equation}
	\begin{split} \label{Eq:Euler_eq}
	\frac{\partial \textbf{v}_{t}}{\partial t}=\frac{e\chi_t}{m}\textbf{E}_{t}-\frac{\textbf{v}_{t}}{\tau_p}-\frac{2 n_b}{n_t + n_b}\frac{\textbf{v}_{t}-\textbf{v}_{b}}{\tau_{D}};
	\\
	\frac{\partial\textbf{v}_{b}}{\partial t}=\frac{e\chi_b}{m}\textbf{E}_{b}-\frac{\textbf{v}_{b}}{\tau_p}-\frac{2 n_t}{n_t + n_b}\frac{\textbf{v}_{b}-\textbf{v}_{t}}{\tau_{D}}.
	\end{split}
	\end{equation}
	where indices $t,b$ denote top and bottom layers, $\textbf{v}_{t},\textbf{v}_{b}$ are to the drift velocities in the upper and lower layers respectively, $n_t$ and $n_b$ are the carrier densities, $e>0$ is the elementary charge, $\chi = +1$ for $p-$doped layer and $-1$ for $n-$doped layer, $m$ is the effective mass (assumed the same for both layers), $\tau_p$ is the effective momentum relaxation time.  The last term of each equation represents the Coulomb drag, $\tau_{D}$ being the characteristic interlayer Coulomb scattering time. It acquires a particularly simple form for equal carrier densities in both layers $n_t = n_b$, and becomes simply $(\textbf{v}_{t}-\textbf{v}_{b})/\tau_D$. It equalizes velocities in the different layers exponentially with time. For dissimilar electron densities $n_t \neq n_b$, the drag acceleration is strong for minority carriers and weak for majority ones. The particular form of density prefactors $n_{t/b}/(n_t + n_b)$ is consistent with total momentum conservation in the double layer the upon interlayer Coulomb scattering.}

	In order to describe plasmons in the system under consideration we apply a standard linearization procedure assuming that the plasmon wave induces a small harmonic in time $e^{-i\omega t + i q_x x}$ variation of charge density, electric current and electric field in the 2DESs. In this case, the dynamic equations~(\ref{Eq:Euler_eq}) can be cast in the form of the 'resistivity matrix':
\begin{equation}
\label{eq-resistance-matrix}
\left( \begin{matrix}
   \rho _{t}^{*} & {{\rho }_{D}}  \\
   {{\rho }_{D}} & \rho _{b}^{*}  \\
\end{matrix} \right)\left( \begin{matrix}
  {{j}_{t}} \\ 
  {{j}_{b}} \\ 
\end{matrix} \right)=
\left( \begin{matrix}
{{E}_{t}} \\ 
{{E}_{b}} \\ 
\end{matrix} \right)
\end{equation}
where $\rho _{t/b}^{*}$ are the resistivities of individual layers renormalized by the drag, and $\rho_D$ is the drag resistivity. They are given by
\begin{gather}
\rho _{t/b}^{*}=\frac{-i\omega +1/{{\tau }_{p}}+1/{{\tau }_{D}}}{{{n}_{t/b}}{{e}^{2}}/m}, \\
{{\rho }_{D}}=\frac{{{\chi }_{t}}}{{{\chi }_{b}}}\frac{1/{{\tau }_{D}}}{\left( {{n}_{t}}+{{n}_{b}} \right){{e}^{2}}/\left( 2m \right)}
\end{gather}
It is instructive to note that the drag resistivity depends on the mutual polarity of carriers in two layers via a factor  $\chi_t/\chi_b$. The symmetry of the resistivity matrix $\hat{\rho}$ is a consequence of total momentum conservation. Another important property of $\hat{\rho}$ is its independence on wave vector, i.e. locality of current-field response. Of course, the latter is just a consequence of neglecting the viscous forces and pressure gradients in the dynamic equations.
	
The dynamic equations are supplemented by the Poisson's equation relating the electric potential $\phi$ with charge densities in the layers $Q_{t/b}$
	\begin{equation}\label{eq-Poisson}
	\Delta\phi=-\frac{4\pi}{\kappa}[Q_{t}(x)\delta(z-z_{t})+Q_{b}(x)\delta(z-z_{b})],
	\end{equation}
where $\kappa$ is the background dielectric constant. Our set of equations is closed by the continuity equation in each of the layers:
\begin{equation}\label{eq-continuity}
	i\omega Q_{t,b} = \partial_x j_{t,b}.
\end{equation}
	
The solution of (\ref{eq-Poisson}) and (\ref{eq-continuity}) allows us to obtain another relation between currents $j_{t/b}$ and in-plane electric field $E = -\partial_x \phi$:
\begin{equation}
\label{eq-field-solution}
    E \left( z \right)=-\frac{2\pi i}{\kappa\omega /q}\left[ {{j}_{t}} e^{-|q_x||z-z_t|}+{{j}_{b}} e^{-|q_x||z-z_b|} \right]
\end{equation}	

Combining the two current-field relations (\ref{eq-resistance-matrix}) and (\ref{eq-field-solution}), and introducing the characteristic impedance 
\begin{equation}
    Z^* = -\frac{2\pi i}{\kappa\omega /q},
\end{equation}
we present the governing equation for the double-layer plasmons in the form
\begin{gather}
\hat{M}\left( \begin{matrix}
  {{j}_{t}} \\ 
 {{j}_{b}} \\ 
\end{matrix} \right)=\left( \begin{matrix}
 0 \\ 
 0 \\ 
\end{matrix} \right), \\ 
\hat{M}=\frac{1}{{{Z}^{*}}}\left( \begin{matrix}
   \rho _{t}^{*} & {{\rho }_{D}}  \\
   {{\rho }_{D}} & \rho _{b}^{*}  \\
\end{matrix} \right)-\left( \begin{matrix}
   1 & {{e}^{-qd}}  \\
   {{e}^{-qd}} & 1  \\
\end{matrix} \right) 
\end{gather}

Equation $\det\hat{M} = 0$ governs the plasmon spectrum, which can be either the complex-valued frequency $\omega$ on real-valued wave vector $q_x$, or vice versa, depending on the wave excitation conditions in a given experiment. In order to visualize these modes, it is convenient to plot the loss function $\mathrm{Im} (1/\det{\hat{M}})$ that incorporates information about resonant frequency $\mathrm{Re}\,\omega$ and linewidth $\propto\mathrm{Im}\,\omega$ simultaneously~\cite{hwang2009plasmon}, see Fig.~\ref{fig:Loss_function}, central line.

Fig.~\ref{fig:Loss_function} consists of three columns. The central column shows the loss function in the absence of Coulomb drag. In this case, the lower out-of-phase acoustic and upper in-phase optical modes are well-defined by their current distributions and their properties do not depend on the doping type in the layers. However, if the layers are coupled via Coulomb drag, only one of the modes experiences extra damping due to an additional friction. To have this extra drag damping, such a mode should possess an asymmetric velocity distribution (see side columns of Fig.~\ref{fig:Loss_function}). The velocity distribution in the $p$/$p$ doping case (right column) coincides with the current distribution (compare the schemes of $j(x)$ in the central column with $v(x)$ in the right column), and acoustic mode becomes overdamped. However, if we switch the carrier type in one of the layers, say, in the bottom layer ($p$/$n$ doping case, left column), the corresponding velocity distribution becomes reversed, and the optical mode acquires asymmetric velocity distribution and thus it becomes overdamped. At the same time the acoustic mode possesses symmetrical velocity distribution and does not experience additional damping.

\refa{To gain further insights into the magnitude of drag-induced damping and its dependence on mutual layer polarity, we present an analytical solution for plasmon dispersion in case of equal absolute values of carrier densities in the layers. In such a case $n_t = n_b = n_0$, $\rho _{t}^{*}=\rho _{b}^{*}=\rho _{L}^{*}$, and the dispersion relation is split into a pair of equations, one for symmetric (optical) mode and the other one for anti-symmetric (acoustic) mode:
\begin{equation}
\rho _{L}^{*}\pm {{\rho }_{D}}={{Z}^{*}}\left( 1 \pm {{e}^{-qd}} \right),
\end{equation}
the upper sign corresponds to the optical mode, and the lower sign -- to the acoustic mode. Plugging the explicit expressions for resistivities, we find
\begin{equation}
\label{eq-dispersion-explicit}
    \omega \left[\omega + \frac{i}{\tau_p} +\frac{i}{\tau_D}\left( 1 \mp \frac{\chi_t}{\chi_b}\right)\right] = \frac{2\pi n_0 e^2 |q_x| (1\pm e^{-|q_x|d})}{\kappa m}.
\end{equation}
From dispersion (\ref{eq-dispersion-explicit}) it becomes clear that inverse Coulomb drag time $1/\tau_D$ selectively contributes to the plasmon damping. For optical modes (upper sign), the effect of drag is finite for electron-hole doping, and is nullified for identical (hole-hole or electron-electron) doping. For acoustic modes (lower sign), the situation is inverted, and the drag contributes to damping for equal charge polarities in the layers.}

\begin{figure}[b]\label{fig:Scheme-confined}
	   \includegraphics[width=0.5\textwidth]{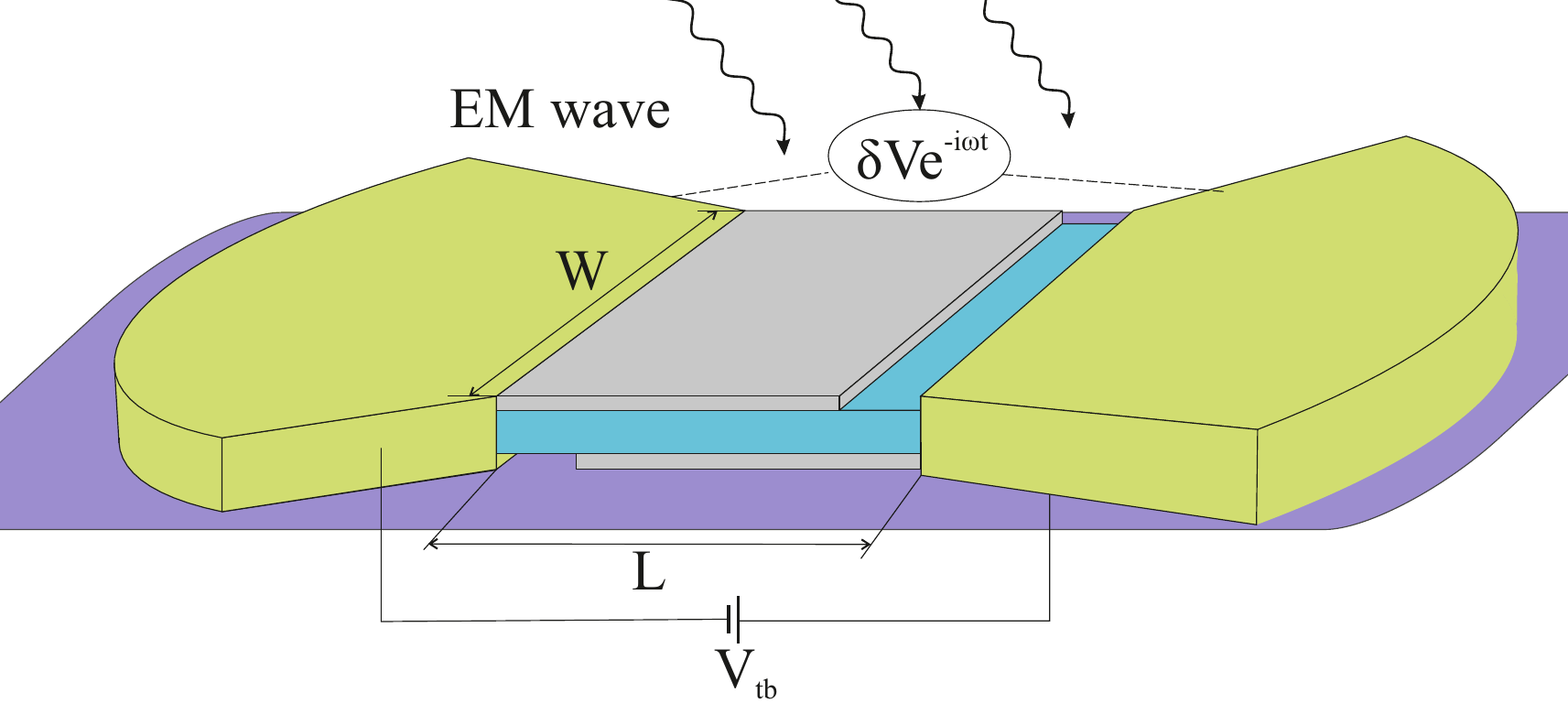}
	   \caption{ Schematic picture of a confined DLR with an antenna irradiated by electromagnetic wave. Thin gray plates denote 2DESs, the spacer is shown in cyan. The 2DESs are biased by the top-bottom voltage $V_{tb}$, which leads to their doping with charge densities $\rho_{t0} = -\rho_{b0}$. The EM wave in converted to the oscillating longitudinal voltage $\delta Ve^{-i\omega t}$ that excites plasmon modes in the DLR. }
\end{figure}

\refa{The effect of drag on plasmon dispersion given by Eq.~(\ref{eq-dispersion-explicit}) is very different from another effect of electron-electron scattering, the electron viscosity. Account of the latter can be achieved (at least for symmetric doping) by amendment of extra term $i \nu q^2$ in the square bracket, where $\nu$ is the kinematic viscosity coefficient. The effects of viscosity appear quadratic in wave vector, while the Coulomb drag appears $q$-independent. It is possible to estimate the range of plasmon wavelengths $\lambda = 2\pi/q$ where the drag-induced damping overwhelms the viscous damping:
\begin{equation}
\lambda \gg 2\pi \sqrt{\nu \tau_D}.
\end{equation}
Explication of this estimate is achieved by considering the fact that the viscosity itself is frequency-dependent,
\begin{equation}
\nu=\frac{1}{4}\frac{v_{0}^{2}\tau_{ee}}{1+(\omega\tau_{ee})^{2}},
\end{equation}
where $v_0$ is the carriers' Fermi velocity and $\tau_{ee}$ is the characteristic time between electron-electron collisions {\it inside of a single layer}. In the low-frequency domain $\omega\tau_{ee} \ll 1$, the dominance of Coulomb drag damping is achieved at wavelengths
\begin{equation}
\lambda|_{\omega\tau_{ee} \ll 1} \gg \pi v_0 \sqrt{\tau_{ee}\tau_D}.
\end{equation}
At high frequencies, the drag-induced damping dominates over the viscous one if
\begin{equation}
\frac{v_0}{2s} \sqrt{\frac{\tau_D}{\tau_{ee}}} \ll 1,
\end{equation}
where we have introduced the phase velocity of the wave $s = \omega/q$. The latter condition is quite non-trivial. From one hand, the speed of 2d plasmons exceeds the Fermi velocity $2s/v_0 > 1$. From the other hand, $\tau_D > \tau_{ee}$ due to the spatial separation between layers. Thus, the drag-induced damping dominates over the viscous damping for relatively fast waves. Of course, all above estimates assumed that the direct drag-induced damping $i/\tau_D$ is allowed by the mode symmetry. In other cases, such as for optical modes in $nn$- or $pp$-doped bilayers, the drag-induced damping is absent at all, and we are left with viscous damping and damping due to phonons and impurities.}

\section{Coulomb drag and plasmons in confined Double-Layer Resonators}\label{SecIII}
	
The previous Section dealt with a model case of infinite DLRs, which means that the plasmon propagation length is much smaller than the dimensions of the sample. The realistic double-layer devices based on graphene separated by dielectric barriers have few-micron size~\cite{Yadav2016,Double_layer_detector1,Ma2016}, which is comparable to the wavelength of 2d plasmons at terahertz frequencies. For such structure dimensions, reflection of 2d plasmons at edges and contacts can strongly affect both resonant frequencies and field distributions. The aim of this Section is to demonstrate that the Coulomb drag effect on plasmons persists for confined DLRs, although the classification of modes according to their symmetry is no more applicable.
	\begin{figure}
		\includegraphics[width=0.5\textwidth]{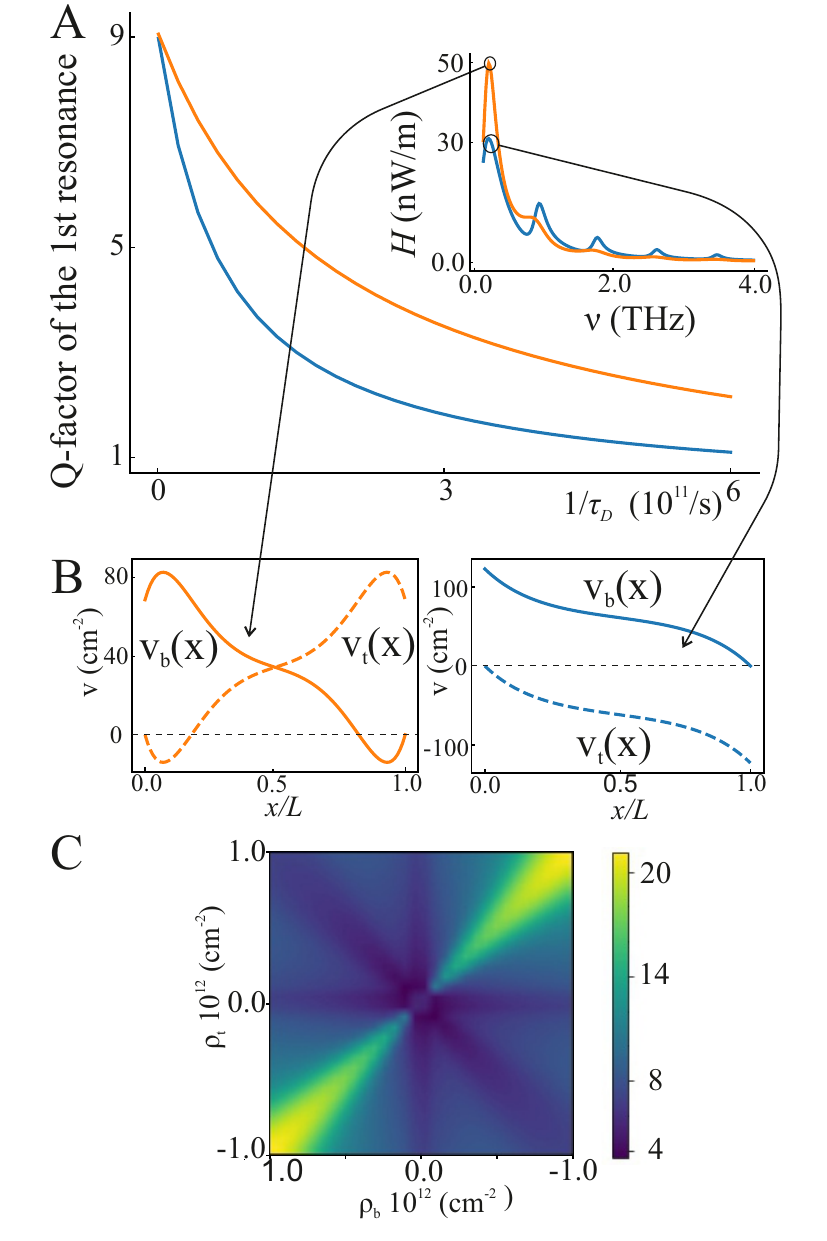}
		\caption{\label{fig-Confined_Plots} (A) Quality factor of the first resonance in confined DLR as a function of Coulomb drag rate at $p$/$p$ doping (blue curve) and $p/n$ doping (orange curve). Note that at zero drag rate the quality factors coincide. Inset: released Joule heat as a function of excitation frequency applied to the antenna. (B) Charge carrier velocity distributions in confined DLR of panel A at the 1st resonance frequency for $p$/$p$ doping (left plot) and $p/n$ doping (right plot). The distribution crucially depends on the doping type of the layers. (C) Color map of the first resonance quality factor as a function of charge density in each of the 2DESs at non-zero Coulomb drag rate. The plot is symmetrical only upon simultaneous reversal of charge carrier type in each of the layers. \textit{Parameters of calculation:}  $L=1$\,$\mu$m, $d=2$\,nm; $|\rho_{t,b}|=10^{16} m^{-2} $ for panels A and B; $ \tau_{p}=20\,\mathrm{ps}$, $\tau_{D}=\tau_p$.}
    \end{figure}

The structure under study is shown in Fig.~\ref{fig:Scheme-confined}. Two parallel layers of 2DES, each of length $L$, are separated by a thin dielectric and are connected to metal source and drain contacts. The top layer is connected to source pad on the left, while its right edge is terminated, and vice versa for the bottom layer. Illumination of the structure results in the buildup of ac voltage $\delta V e^{-i\omega t}$ between source and drain contacts. Due to the plasmonic effects, the voltage profiles along the layers $\phi_{t/b}(x)$ are highly non-uniform, and the local voltage difference $\phi_{t}(x)-\phi_{b}(x)$ can largely exceed the input voltage $\delta V$ at the resonant condition. Our next goal will be to study the properties of plasmon resonance in such a structure, and reveal the effect of Coulomb drag on these resonances.

Our theoretical approach is based on the gradual channel approximation which states that for smoothly varying potential profiles the Poisson equation (\ref{eq-Poisson}) can be replaced by the plane capacitor model
	\begin{equation}\label{eq-plane=capacitor-model}
	\begin{split}
		& \rho_{t}(x) + \rho_{b}(x) = 0; \\
		& C(\rho_{t}(x)-\rho_{b}(x))=\phi_{t}(x)-\phi_{b}(x),
	\end{split}
	\end{equation}
	where $C = \kappa/8\pi d$, $\kappa$ is the spacer permittivity, $d$ is the interlayer distance. The first equation is the charge conservation law, which presumes the asymmetric charge distribution. Correspondingly, our model does not describe the optical mode. Still, the consideration of the acoustic mode solely is enough for the purposes of this Section.
	
	We make use of Eqs.\,(\ref{eq-plane=capacitor-model}), the resistivity tensor (\ref{eq-resistance-matrix}), and the continuity equation, thus arriving at	
	\begin{equation}\label{eq-confined}
	-i\omega\begin{pmatrix}
		\chi_{t}\dfrac{\phi_t - \phi_b}{2C} \\
		-\chi_{b}\dfrac{\phi_t - \phi_b}{2C}
	\end{pmatrix} =
	\hat{\rho}^{-1} \begin{pmatrix}
	\partial_{xx}\phi_t \\
	\partial_{xx}\phi_b
	\end{pmatrix}.
	\end{equation}

We assume the standard boundary conditions for confined DLR, namely the plasmon wave potential difference between the contacts 2DESs/antenna is equal to the antenna voltage, and the normal current at the 2DES edge must vanish:
	\begin{equation}\label{eq-confined-BCs}
	\begin{split}
	j_{b}|_{x=0}=0 ;\phi_{t}|_{x=0}=\frac{\delta V}{2};\\   
	\phi_{b}|_{x=l}=-\frac{\delta V}{2};j_{t}|_{x=l}=0.
	\end{split}
	\end{equation}
	
The problem (\ref{eq-confined}), (\ref{eq-confined-BCs}) incorporates an external "force" acting on the DLR, which is the antenna voltage (ususally, several $\mu$V~\cite{mylnikov2022terahertz,titova2023ultralow}). Consequently, we no longer need to plot the loss function to visualize plasmon modes; instead we evaluate the Joule heat released in the system
	\begin{equation}
		H = \langle(\mathbf{j}(t,x),\mathbf{E}(t,x))\rangle,
	\end{equation}
where the brackets indicate integrating over the device length and time averaging. A typical $H(\omega)$ dependence (Fig.\ref{fig-Confined_Plots}A, inset) exhibits several peaks that correspond to spatial resonances, and in the presence of Coulomb drag the peaks widen, and their quality factor crucially depend on the layer doping type.

Naturally, in the absence of Coulomb drag ($1/\tau_D = 0$) the quality factor $Q$ of these resonances is independent of the doping type of the 2DESs (Fig.~\ref{fig-Confined_Plots}A), and it decreases with different slopes for differently doped 2DESs. This distinction stems from disparate charge carrier velocity profiles of the acoustic plasmon for $p$/$p$ doping and $p$/$n$ doping (Fig.~\ref{fig-Confined_Plots}B). Fig.~\ref{fig-Confined_Plots}C provides a wider outlook on the Q-factor dependence on the doping type and doping level in the 2DES at finite Coulomb drag time; switching off the Coulomb drag would make the chart symmetric under reflection relative to any of the axes. 

Thus, in confined systems the Coulomb drag directly affects the spatial profiles of plasmon modes which makes its analysis much more intricate. Still, the main conclusion remains the same: the Coulomb drag effect is a tool that allows to distinguish multilayer 2DESs with different doping types.

\section{Discussion and conclusion}
		
In conclusion, we have shown that long-range Coulomb interactions affect the plasmon modes in multilayer systems by providing an additional damping mechanism. This influence is local and, moreover, is most pronounced at low wave vectors and frequencies comparable with inverse drag time $\omega \lesssim \tau_D^{-1}$. The lowering of quality factor crucially depends on the doping type of the layers and may be totally absent for modes and doping types such that drift velocities of charge carriers in two layers are co-directional. We stress that the predicted effect cannot be predicted from macroscopic electrodynamics calculations, where the current density in 2DES is proportional to its conductivity. Indeed, within such treatment, the layers with identical carrier densities and effective masses are electrodynamically indistinguishable.

The above discussion was based on hydrodynamic equations of motion for charge carriers in the two layers. It is generally accepted that such equations are applicable for wave frequencies $\omega$ below the carrier-carrier collision frequency within the layer $\tau_{ee}^{-1}$. Yet, several recent exact solutions of the kinetic equation with model carrier-carrier collision integrals have shown that the applicability of hydrodynamics is much broader, though the coefficients of hydrodynamic equations can be renormalized. In the absence of magnetic fields, hydrodynamic formulation is possible provided $qv_0/|\omega+i/\tau_{ee}| \ll 1$~\cite{Svintsov_crossover}, while in finite magnetic fields the criterion is $q v_0/\omega_c \ll 1$, where $\omega_c$ is the cyclotron frequency~\cite{kapralov2022ballistic,Alekseev_viscoelastic_resonance,afanasiev2023new}. 

Finally, we would like to provide numerical estimates of 2DES parameters where the predicted effects can be observable. The drag time can be estimated as a quasiparticle lifetime due to electron-electron scattering~\cite{Li_Das-Sarma_free_path,zheng1996coulomb} timed by $\exp(2qd)$, where $q$ is the characteristic momentum transferred upon collisions. Taking for estimate $q \sim k_F$, where $k_F$ is the Fermi momentum, for 2DEG with parabolic energy spectrum we arrive at~\cite{zheng1996coulomb}
\refa{
\begin{equation}
    \tau_D^{-1} \simeq \frac{\pi E_F}{8\hbar}\left(\frac{kT}{E_F}\right)^2\ln\left(\frac{E_F}{kT}\right)\times \exp(-2 k_F d).
\end{equation}
}
The resulting drag time for $n_t=n_b=10^{12}\,$cm$^{-2}$, $m=0.03\,m_e$ (bilayer graphene) and realistic interlayer distance $d=10$ nm equals $0.1$ ps at $T=300$ K, 0.5 ps at $T=77$ K, and 100 ps at $T=4$ K. While at lowest considered temperatures $\tau_D^{-1}$ will be definitely overwhelmed by impurity scattering rate, the effects of drag on plasmon damping already at $T\sim 77$ K can be pronounced. Anyway, these effects can be enhanced for ultra-proximized double-layer structures, such as graphene flakes separated by few-layer boron nitride dielectrics. The only possible to downscaling of interlayer separation is the interlayer tunneling. The latter would affect the properties of plasmons at frequencies $\omega < t/\hbar$, where $t$ is the tunnel splitting between 2DES~\cite{DasSarma1998}. According to the microscopic estimates~\cite{alymov2023ultimate}, the tunneling rate for few-layer boron nitride dielectrics is as low as $t\sim 1$ meV. Thus, the effects of Coulomb drag on plasmon lifetime in double layers should be observable in a broad range of dielectric thicknesses and at frequencies around units of terahertz.
	 
\section{Acknowledgement}
The work was supported by the grant 22-29-01034 of the Russian Science Foundation. I.S. thank Vladimir Potanin via the Center for Neurophysics and Neuromorphic Technologies.

\bibliography{apssamp}

\end{document}